\documentclass[11pt,a4paper,onecolumn]{article}
\usepackage[utf8]{inputenc}
\usepackage{authblk}
\usepackage[utf8]{inputenc}
\usepackage[english]{babel}
\usepackage{amsmath}
\usepackage{subfig}
\usepackage{braket}
\usepackage{relsize}
\usepackage{amsfonts}
\usepackage{amssymb}
\usepackage{graphicx}
\usepackage[left=2cm,right=2cm,top=2cm,bottom=2cm]{geometry}
\usepackage{cite}
\usepackage{ifplatform}
\usepackage{amsmath,amssymb,amsfonts}
\usepackage{algorithmic}
\usepackage{graphicx}
\usepackage{textcomp}
\usepackage{listings}
\usepackage{fancyvrb}
\usepackage{framed}
\usepackage[listings,skins]{tcolorbox}
\usepackage{xcolor}
\usepackage[utf8]{inputenc}
\usepackage[english]{babel}
\usepackage{float} 
\usepackage{hyperref}
\usepackage{accsupp}

\usepackage[skipbelow=\topskip,skipabove=\topskip]{mdframed}
\def\BibTeX{{\rm B\kern-.05em{\sc i\kern-.025em b}\kern-.08em
    T\kern-.1667em\lower.7ex\hbox{E}\kern-.125emX}}

\providecommand{\keywords}[1]
{
  \small	
  \textbf{\textit{Keywords---}} #1
}
  
\title{Thermal relaxation error on QKD: Effect and A Probable Bypass}
\author[1]{Munsi Afif Aziz\thanks{{munsiafifaziz@gmail.com}}}
\author[1]{Bishwajit Prasad Gond\thanks{{bishwajitprasadgond@gmail.com}}}
\author[1]{Srijita Nandi\thanks{{srijitanandi22@gmail.com}}}
\author[2]{Soujanya Ray\thanks{{shaun.ray.1996@gmail.com}}}
\author[3]{Debasmita Bhoumik}
\author[3]{Ritajit Majumdar}
\affil[1]{\textit{Department of Computer Science \& Engineering\\
Government College of Engineering and Leather Technology, India}}
\affil[2]{\textit{National Institute of Technology, Sikkim}}
\affil[3]{\textit{Advanced Computing \& Microelectronics Unit,
Indian Statistical Institute, Kolkata, India}}

\begin{document}
\date{}
\maketitle


\begin{abstract}
    Quantum cryptography was proposed as a counter to the capacity of quantum computers to break classical cryptosystems. A broad subclass of quantum cryptography, called quantum key distribution (QKD), relies on quantum mechanical process for secure distribution of the keys. Quantum channels are inherently noisy, and therefore these protocols will be susceptible to noise as well. In this paper, we study the performance of two QKD protocols - BB84 and E91 under thermal relaxation error. We show that while E91 protocol loses its security immediately due to loss of entanglement, the performance of BB84 protocol reduces to random guessing with increasing time. Next, we consider the action of an Eve on the BB84 protocol under thermal relaxation noise, who is restricted to guessing the outcome of the protocol only. Under this restriction, we show that Eve can still do better than random guessing when equipped with the characteristics of the noisy channel. Finally, we propose a modification of the BB84 protocol which retains the security of the original protocol, but ensures that Eve cannot get any advantage in guessing the outcome, even with a complete channel information.
\end{abstract}



\section{Introduction}

Cryptography nowadays is one of the most important aspects of our security systems. Classical cryptography, which is broadly used in our day-to-day lives and can be classified into two types - symmetric or private key cryptography and, asymmetric or public key cryptography - is based on computational hardness of some problems such as prime factorization and discrete logarithmic. However, Shor \cite{shor1994algorithms} proposed a polynomial time quantum algorithm for both of the above mentioned problem, thus rendering classical cryptography vulnerable in a quantum world.

To counter this probable vulnerability, cryptography using principles of quantum mechanics, such as Uncertainty Principle \cite{heisenberg1985anschaulichen} and the No Cloning theorem \cite{wootters1982single}, called quantum cryptography, has been proposed. There are several branches of quantum cryptography such as Quantum Key Distribution \cite{shor2000simple}, Quantum Secret Sharing \cite{hillery1999quantum}, Quantum Secure Direct Communication \cite{das2021quantum}, Quantum Authentication \cite{majumdar2021sok, dutta2021short}, to name a few. In this study, we shall restrict ourselves to Quantum key distribution (QKD), which deals with sharing a secret random key between two parties, Alice and Bob, which then can be used for future communication. The theory of quantum mechanics posits that the state of an arbitrary qubit is changed upon observation. Hence, the presence of an eavesdropper in the channel can be confirmed with high probability by the user at the other end by measuring the qubits and comparing a fraction of the outcomes with the sender. If no presence of eavesdropper is observed, then the channel is declared to be secure from eavesdropper.

However, quantum channels are, in general, noisy. Therefore, any deviation in the performance of the QKD protocol may be due to the presence of an Eavesdropper, Eve, or due to channel noise, or both. Eve can hide herself behind the noise as long as her interaction with the system can be disguised as the effect of noise. A question, then, is whether it is possible for Eve to gain any beneficial information while still hiding behind noise. In this paper, we restrict ourselves to a thermal relaxation noise channel. In this type of noise, a system spontaneously releases energy and tries to settle in the ground state. Thermal relaxation error is parameterized by $T_1$ such that, if a qubit was prepared in the excited state $\ket{1}$, the probability of finding the state in $\ket{1}$ after time $t$ is $exp(-t/T_1)$. We also restrict the powers of Eve such that: (i) Eve has complete knowledge of the noise in the quantum channel (i.e. the $T_1$ value and the time duration $t$ of the channel), (ii) Using this information, Eve can only \emph{make a guess} about the bitstrings received by Bob after measurement.

Two of the earliest QKD protocols are BB84 \cite{bb84} and E91 \cite{ekert1992quantum}. While the later uses entanglement, the former does not require it. We choose these two as a broad subclass between QKDs as those do and do not require entanglement. Keeping aside the advantages and disadvantages of using entanglement from the view point of any other criteria, we study the effect of thermal relaxation noise on both of these protocols. We analytically and, also through simulation, study the effect of thermal relaxation noise on the success probability of the said protocols, and show that:

\begin{itemize}
    \item the success probability of BB84 protocol converges to $\frac{1}{2}$ as $t \rightarrow \infty$.
    \item the success probability of E91 protocol first reduces, and then again increases to $1$ as $t \rightarrow \infty$.
    \item we argue that even though apparently E91 has perfect fidelity, as $t \rightarrow \infty$, Eve can guess the correct outcome with probability $p \rightarrow 1$ as $t \rightarrow \infty$.
    \item restricting ourselves to the regime of finite $t$, we argue that Eve can guess the correct outcome of the BB84 protocol with probability $1 > p \geq \frac{1}{4} + \epsilon$, for some $\epsilon > 0$.
    \item we propose a simple variation of the BB84 protocol that ensures that even in thermal relaxation channel, Eve does not have any upper-hand when guessing the correct outcome.
\end{itemize}

The rest of the paper is organized as follows - In Sec. 2 we provide a brief overview of the two QKD protocols used and the noise model. Sec. 3 shows the analytical and the simulation results of the two QKD protocols in the presence of thermal relaxation noise. We argue that the probability of success of an Eve, who makes a guess about the final bitstrings given the information of the noise channel, is higher than random guessing. In Sec. 5, we propose a modification of the BB84 protocol such that the protocol works like usual in a noiseless scenario, but Eve is restricted to random guessing even in a thermal relaxation channel. We conclude in Sec. 6.

\section{Background}
In this section we briefly discuss BB84 and E91 protocols, and the Kraus Operator representation of a thermal relaxation channel.

\subsection{Brief description of BB84 protocol }
First developed by Charles Bennett and Gilles Brassard in 1984\cite{bb84} and based on the Quantum Conjugate Coding proposed by Wiesner in the late 1960s\cite{wiesner1983conjugate}, BB84 was the first-ever quantum key distribution protocol to be created. The brief overview of this protocol is as follows:
Alice chooses two uniformly random binary strings $A=a_{1}a_{2}a_{3}...a_{2n}$ and $B=b_{1}b_{2}b_{3}...b_{2n}$ ,$a_{i},b_{i} \in{0,1}$. Alice then encodes her data bits as $\ket{\psi_{a_{i}b_{i}}}$ using four states
    $\ket{0},\ket{1},\ket{+},\ket{-}$. If $a_{i}$ is 0, then the state is either $\ket{0}$ or $\ket{+}$ and, $\ket{1}$ or $\ket{-}$ if $a_{i}$ is 1. The base is decided by $b_{i}$. If $b_{i}$ is 0, the state is either $\ket{0}$ or $\ket{1}$ and, if it is 1, then the state is $\ket{+}$ or $\ket{-}$. This base is then sent to Bob. Bob receives the qubits, acknowledges the receival publicly, and measures each qubit in the X or Z basis at random. Alice and Bob publicly reveal their bases and compare their bases. They discard all the bits where Bob measured his qubits in a different basis than the one Alice prepared. There is a high probability that there are at least $n$ bits left. If that is not the case, they abort the protocol. Alice selects $n/2$ bits to check on Eve's interference and tells Bob through a classical channel. Alice and Bob announce and compare the values of the $n/2$ check bits. If the comparison result  does not show an acceptable number of consistent bits despite having chosen the same basis, they abort the protocol. Alice and Bob perform information reconciliation and privacy amplification on the remaining $n/2$ bits to obtain a secret key. Information Reconciliation allows two parties knowing correlated random variables to agree on a shared string. Privacy amplification is a process that allows two parties to distil a secret key from a common random variable about which an eavesdropper has partial information.
\\

\subsection{Brief description of E91 protocol } 
E91 is also a QKD protocol proposed by Arthur K.Ekert in 1991,\cite{ekert1992quantum} which utilizes the idea of the Bell's Theorem\cite{bell1965physics} and Clauser-Horne-Shimony-Holt(CHSH) or Bell’s Inequality \cite{clauser1969proposed}.The EPR pair is used because of their entanglement properties, which are as follows:
\[ |\phi^{+}\rangle = \frac{1}{\sqrt{2}}|00\rangle + \frac{1}{\sqrt{2}}|11\rangle,|\phi^{-}\rangle = \frac{1}{\sqrt{2}}|00\rangle - \frac{1}{\sqrt{2}}|11\rangle,\]
\[ |\psi^{+}\rangle = \frac{1}{\sqrt{2}}|01\rangle + \frac{1}{\sqrt{2}}|10\rangle, |\psi^{-}\rangle = \frac{1}{\sqrt{2}}|01\rangle - \frac{1}{\sqrt{2}}|10\rangle,\]  

Also, the CHSH inequality is used to confirm the maximum entanglement of the shared qubit pair. Due to monogamy of entanglement i.e., if two parties share a maximally entangled qubits then a third party cannot have any entanglement with it, they use the CHSH inequality to their own favour detecting this. In this protocol, Alice and Bob both have access to a Quantum state generating source and they are connected by a classical channel. The source sends $n$ entangled pair of qubits, one for each entangled state, to Alice and Bob. Alice and Bob choose $A1, A2, A3$ and $B1, B2, B3$ for their measurement of entangled states respectively.\\
\begin{center}
    $A_1 = Z,$  $A_2 = X,$  $A_3 = \frac{1}{\sqrt{2}}(Z+X),$\\
 $B_1 = Z,$ $B_2 =\frac{1}{\sqrt{2}}(Z-X),$ $B_3 = \frac{1}{\sqrt{2}}(Z+X)$ 
\end{center}

 Alice and Bob then announce the measurements basis for each state. They used the basis that matches as the key and used the other states to test the entanglement of their shared states. For a classical set of observable, the Tsirelson’s bound\cite{tsirel1987quantum} for the chosen observable is as follows:
 \begin{center}
     Key Generation:- $(A_1,B_1)$ \& $(A_3,B_3)$

Test: $(A_1,B_3)$, $(A_1,B_2)$, $(A_2,B_3)$ \& $(A_3,B_2)$
 \end{center}

For classical random variable $A_{1}$, $A_{2}$, $B_{3}$, $B_{2}$ with realization $\pm 1$.

\[ A_1(B_{3} + B_{2}) + A_1(B_{3} - B_{2}) = \pm 2\]
\[| \langle A_1(B_{3} + B_{2}) + A_1(B_{3} - B_{2})\rangle| \leqslant 2\]
\[S = | \langle A_1 B_{3}\rangle +  \langle A_1 B_{2}\rangle + \langle A_2 B_{3}\rangle - \langle A_2 B_{2}\rangle| \leqslant 2\]

Assuming $A_{1}, A_{2}, B_{3}, B_{2}$ as Quantum observable
\[\langle A_i B_{j}\rangle = Tr( A_i\otimes B_{j})\]

For E91 protocol:
\[A= Z, B=\frac{1}{\sqrt{2}}(Z+X), \rho= |\psi^- \rangle  \langle\psi^{-} |\]
\[ \therefore A_1 B_3 = \langle \psi^{-}|\: Z \otimes \frac{1}{\sqrt{2}} (Z+X) |\psi^{-}\rangle = \frac{-1}{\sqrt{2}}\]
\[S = | \langle A_1 B_{3}\rangle +  \langle A_1 B_{2}\rangle + \langle A_2 B_{3}\rangle - \langle A_2 B_{2}\rangle| \leqslant 2\sqrt{2}\]

After measuring, if their CHSH value is $2\sqrt{2}$, then their states are maximally entangled because it violates the classical CHSH inequality, if it is between 2 and  $2\sqrt{2}$,  then they use classical post processing to turn this partially correlated and partially secure key into a secure key and, if it is less than or equal to 2, then there is no entanglement left and they discard the whole key.

\subsection{Thermal Relaxation Error}
Thermal relaxation is the phenomenon where a system, prepared in some higher energy state, has a tendency to spontaneously release the energy and settle in the ground state. We call a noisy quantum channel that exhibit such a phenomenon a \emph{thermal relaxation channel}. Such a channel is non unitary, and its evolution can be described by the Kraus Operator representation \cite{nielsen2002quantum}. The evolution of some quantum state $\rho$ in a thermal relaxation channel for some time $t$ can be expressed as:

$$\rho \rightarrow\rho' = \displaystyle \sum_{k=1}^2 E_k \rho E_k^{\dagger},$$

\begin{center}
    $E_1 = \begin{bmatrix}
            1 & 0\\
            0 & \sqrt{1-\lambda}
            \end{bmatrix}$ \quad $E_2 = \begin{bmatrix}
                                        0 & \sqrt{\lambda}\\
                                        0 & 0
                                        \end{bmatrix}$
\end{center}

where, $\lambda =1 - e^{-t/T_1}$. It can be verified from the above expression that, given a state prepared in $\ket{1}$, the probability of measuring a state in $\ket{1}$, after a time $t$, is $e^{-t/T_1}$. Higher the value of $T_1$, slower is the decay. The $T_1$ values of current IBM Quantum hardware is in the range of $100 \mu s$.

\section{Effect of thermal relaxation error on BB84 and E91 protocols}
In this section. we analytically show the effect of thermal relaxation error on the two protocols mentioned before, and verify them via simulations as well. All the simulation results are generated using qiskit sdk \cite{qiskitsdk}. 

\subsection{BB84 Protocol under Thermal Relaxation error:}

For the analysis of this protocol under the effect of noise, we only consider those situations where Alice and Bob select the same basis for preparation and measurement respectively. Scenarios where the basis differ, will anyway be discarded, and hence, need not come under this analysis.

In this protocol, Alice chooses each of the four states randomly, i.e., probability of Alice sending any state among $\ket{0}$, $\ket{1}$, $\ket{+}$, $\ket{-}$ is $\frac{1}{4}$. Let $\rho_i$ be the density matrix representation of the state $i$ for $i \in \{\ket{0}, \ket{1}, \ket{+}, \ket{-}\}$.

Now, let us assume for some state $\rho$, after being exposed to thermal relaxation, error evolves to $\rho'$. The evolution of $\rho$ is given as:

\begin{center}
    $\rho \rightarrow \rho' = \sum_k E_k\rho E_{k}^{\dagger}$
\end{center}

where, $E_i$ are the Kraus operators for the thermal relaxation noise model.

Thus, the evolution of each $\rho$ for some time $t$ is as follows:
\begin{eqnarray*}
\rho'_{0}&=& \begin{bmatrix}
1 & 0\\
0 & 0
\end{bmatrix} \text{, } \rho'_{1} = \begin{bmatrix}
\lambda & 0\\
0 & 1$-$\lambda
\end{bmatrix}\\ \newline
\rho'_{+}  &=& \frac{1}{2} \begin{bmatrix}
1+\lambda & \sqrt{1-\lambda} \\
\sqrt{1-\lambda} & 1$-$\lambda
\end{bmatrix}\\ \newline
\rho'_{-}  &=& \frac{1}{2} \begin{bmatrix}
1+\lambda & -\sqrt{1-\lambda}\\
-\sqrt{1-\lambda} & 1$-$\lambda
\end{bmatrix}
\end{eqnarray*}

\begin{figure}[H]%
    \centering
    \subfloat[\centering Change of Probability of Bob measuring $\ket{1}$ ]{{\includegraphics[width=7.5cm]{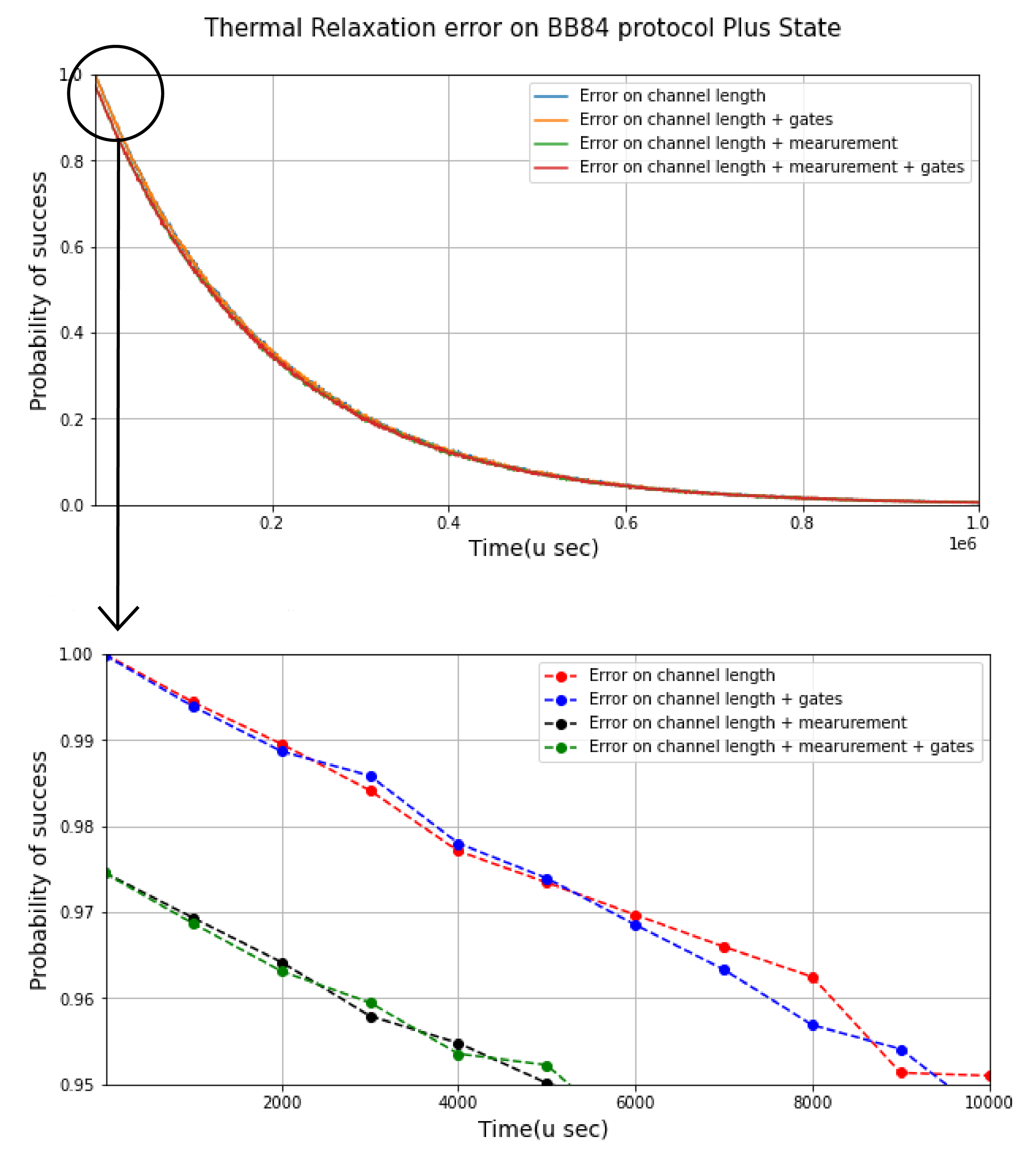}}}%
    \centering
    \qquad
    \subfloat[\centering Change of Probability of Bob measuring $\ket{+}$/$\ket{-}$ ]{{\includegraphics[width=7.5cm]{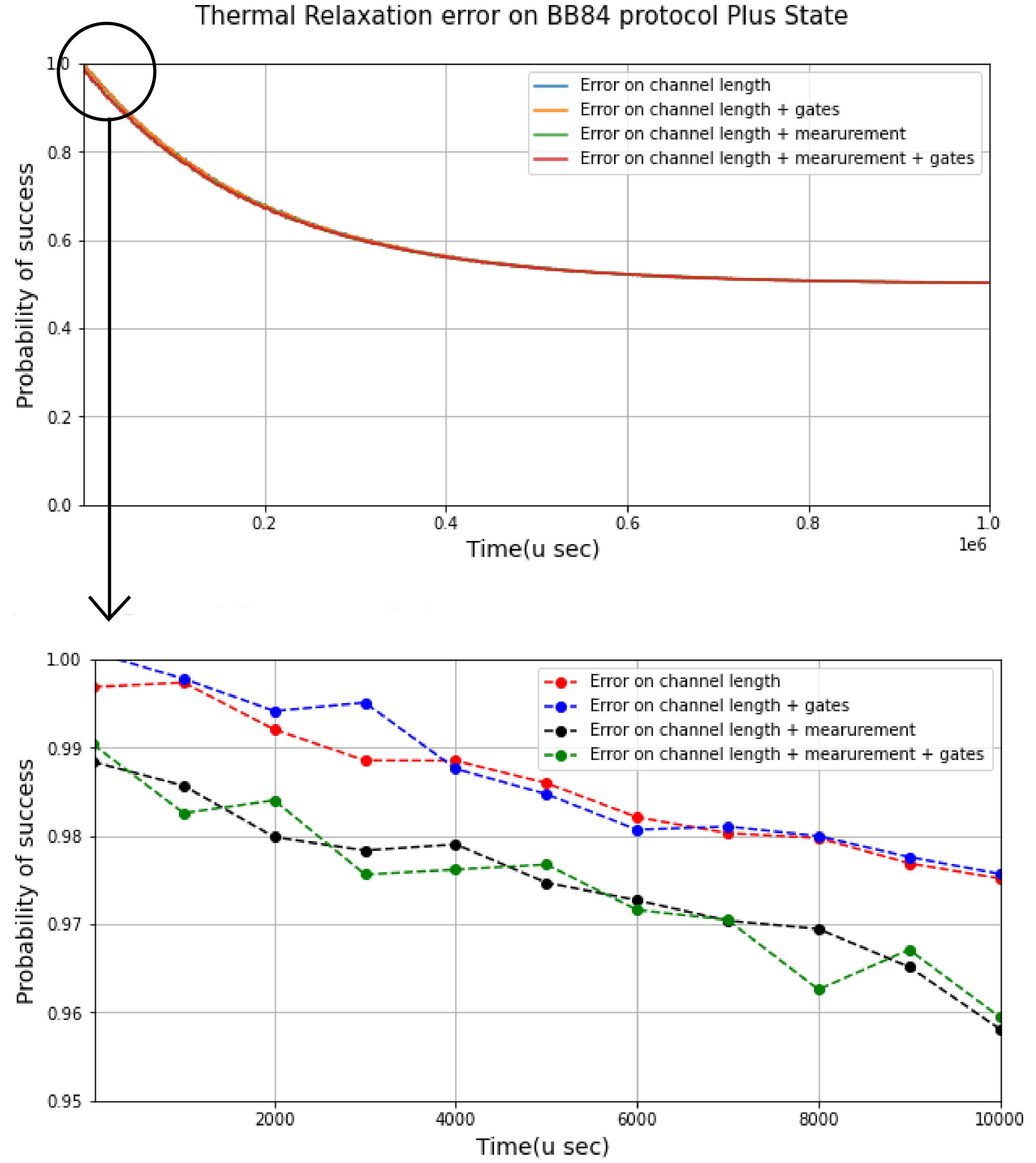} }}%
        \caption{Change of measurement Probability of Hadamard states when measured in Z basis under Thermal Relaxation Error }%
    \label{fig:var1}%
\end{figure}

\newpage So, the probability for correct measurement outcome for each state is:

\begin{center}
    $Tr(\ket{0}\bra{0} \rho'_0)=1$, $Tr(\ket{1}\bra{1} \rho'_{1})=(1-\lambda)=e^{-t/T_{1}}$ \\
    $Tr(\ket{+}\bra{+} \rho'_{+})=\frac{1}{2}(1+\sqrt{1-\lambda})=\frac{1}{2}(1+e^{-t/2T_{1}})$\\
    $Tr(\ket{-}\bra{-} \rho'_{-})=\frac{1}{2}(1+\sqrt{1-\lambda})=\frac{1}{2}(1+e^{-t/2T_{1}})$
    \end{center}
\begin{figure}[H]%
    \centering
    \subfloat[\centering Change of Probability of measurement $\ket{0}$ and $\ket{1}$ when $\ket{+}$ is measured in Z basis ]{{\includegraphics[width=7.5cm]{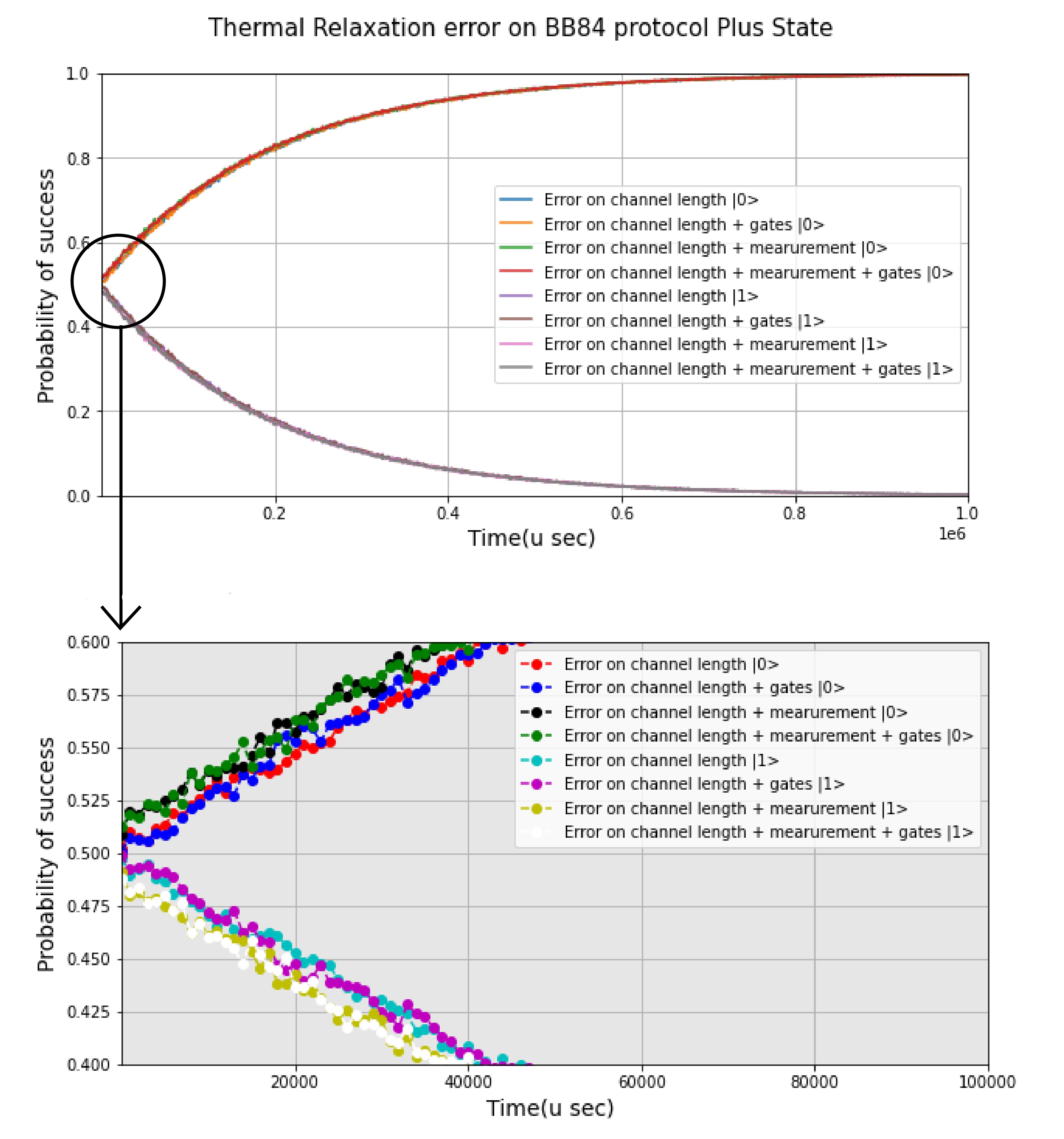} }}%
    \centering
    \qquad
    \subfloat[\centering Change of Probability of $\ket{0}$ and $\ket{1}$ when $\ket{-}$ is measured in Z basis ]{{\includegraphics[width=7.5cm]{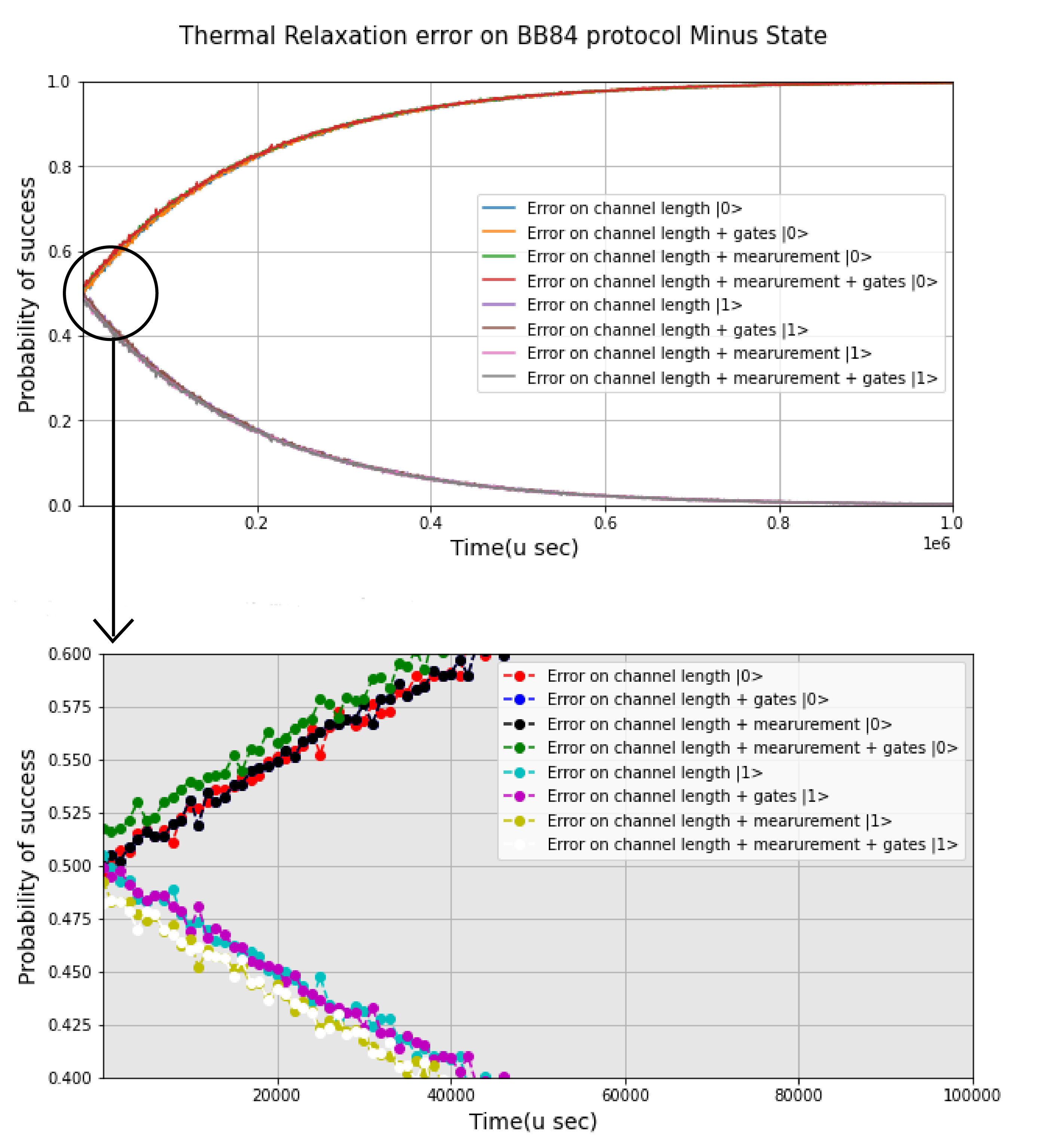} }}%
    \caption{Change of measurement Probability of measurement Hadamard states when measured in Z basis under Thermal Relaxation Error }%
    \label{fig:var2}%
\end{figure}
Therefore, the probability of correct outcome for Bob is now time dependent. When the state is $\ket{0}$, the measurement outcome remains constant, since it is the ground state, but the state $\ket{1}$ decays with increasing time. The measurement probability for both $\rho_+$ and $\rho_{-}$ are decaying as well, and as $t \rightarrow \infty$, both the states collapse to $\ket{0}$. Therefore, the measurement outcomes in the $X$ basis for both of these states approach $\frac{1}{2}$. The variation in the probability when measuring $\ket{1}$, $\ket{+}$ and $\ket{-}$ are shown in Fig.~\ref{fig:var1} and Fig.~\ref{fig:var2} respectively. Let the probability of measuring a state $i$ by Bob, given the same state was sent by Alice be $ P(B=i|A=i)$, and $P(A=i)$ be the probability that Alice chooses to send the state $i$. For this particular protocol, $P(A=i) =\frac{1}{4}$, $\forall$ $i \in \{\ket{0},\ket{1},\ket{+},\ket{-}\}$. Here,
\begin{eqnarray*}
    P(B=\ket{0}|A=\ket{0}) &=& Tr(\ket{0}\bra{0} \rho'_0)\\
    P(B=\ket{1}|A=\ket{1}) &=& Tr(\ket{1}\bra{1} \rho'_1) \\
    P(B=\ket{+}|A=\ket{+}) &=& Tr(\ket{+}\bra{+} \rho'_{+}) \\
    P(B=\ket{-}|A=\ket{-}) &=& Tr(\ket{-}\bra{-} \rho'_{-}) 
\end{eqnarray*}
The protocol is successful if Bob obtains the same state that was sent by Alice. Therefore, the probability of success is:
\begin{eqnarray*}
    P(success) &= \displaystyle \sum_i P(B=i|A=i).P(A=i)
\end{eqnarray*}

\begin{center}
    $=\frac{1}{4}(1) + \frac{1}{4}e^{-t/T_{1}} +\frac{1}{4}(\frac{1}{2}(1+e^{-t/T_{1}}))\times 2$\\
$=\frac{1}{4}(2+e^{-t/T_1}+\sqrt{e^{-t/T_1}})$\\
\end{center}

\begin{figure}[htp]%
    \centering
    {{\includegraphics[width=7.5cm]{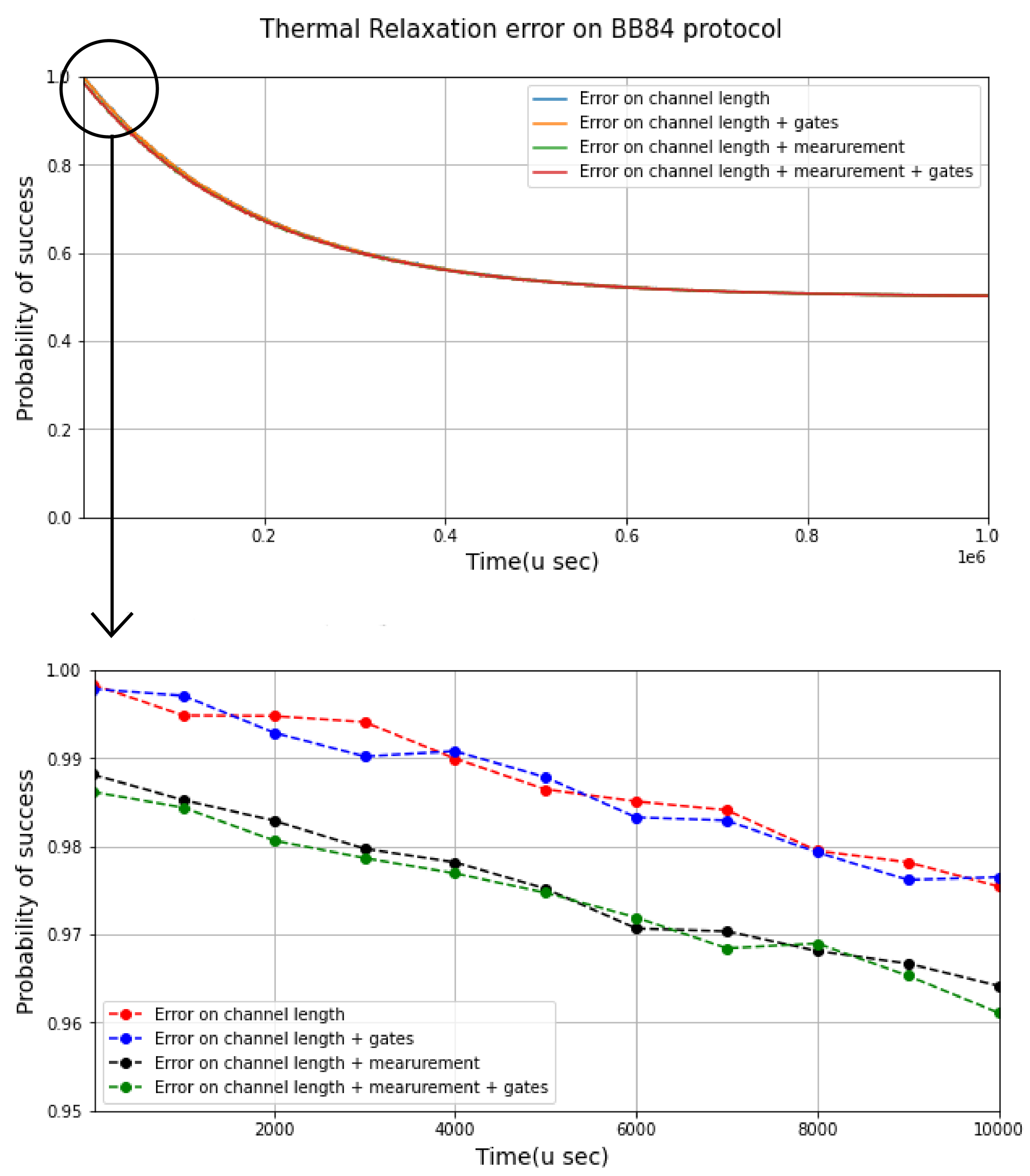} }}%
    \caption{ Probability of Success}%
    \label{fig:var3}%
\end{figure}

With increasing time, the probability of success decreases, and as time $t \rightarrow \infty$ the success probability converges to $\frac{1}{2}$. Therefore, at this point, the protocol is nothing better than random guessing.

In Fig.~\ref{fig:var3},  we verify this using the IBM Quantum simulator Qiskit \cite{qiskit}. From the Fig.~\ref{fig:var1}.a, the scenario of the state $\ket{1}$ is very much evident, with increasing time the state decays to 0. For the states $\ket{+}$ and $\ket{-}$, in Fig.~\ref{fig:var1}.b the plot shows the decay of Hadamard states and after certain time, it saturates at $\frac{1}{2}$, creating a equilibrium of both the Hadamard states with equal probabilities. Since we are restricted by qiskit to perform measurement in Z basis only, the Fig.~\ref{fig:var2} looks into the Hadamard states to confirm that the $\ket{1}$ state is decaying to $\ket{0}$, increasing its outcome probability.

\subsection{E91 Protocol under Thermal Relaxation Error:}
Similar to our analysis of the BB84 protocol, in this section, we analytically show the performance of E91 protocol under thermal relaxation error. We analyze two scenarios corresponding to the underlying quantum channel being $\ket{\phi^{+}}$ and $\ket{\psi^{+}}$. As before, we define \emph{success} to be the scenario where Alice and Bob obtain the same states. Therefore, for an entangled state, there are two different ways such that Alice and Bob end up with the same outcomes (Bob may require to apply a necessary Pauli operator according to the quantum channel).

\subsubsection{E91 protocol with $\ket{\phi^{+}}$ as the quantum channel}
Let us assume the quantum channel is:
\begin{eqnarray*}
 \ket{\phi^{+}} &=&\frac{1}{\sqrt{2}}(\ket{00}+\ket{11})\\
 \therefore \rho = \ket{\phi^{+}}\bra{\phi^{+}} &=& \frac{1}{2}\begin{bmatrix}
1 & 0 & 0 & 1\\
0 & 0 & 0 & 0\\
0 & 0 & 0 & 0\\
1 & 0 & 0 & 1
\end{bmatrix}  \\
\end{eqnarray*}
The evolution of this state under thermal relaxation noise is:
$ \rho \rightarrow \rho' = \sum_{i,j}E_{ij}\rho E_{ij} ^\dagger$, where $E_{ij}=E_i \otimes E_j$  


\begin{center}

= $\frac{1}{2}\begin{bmatrix}
(1+\lambda)^2 & 0 & 0 & 1-\lambda\\
0 & \lambda(1-\lambda) & 0 & 0\\
0 & 0 & \lambda(1-\lambda)  & 0\\
1-\lambda & 0 & 0 & (1-\lambda)^2
\end{bmatrix}$ \\
    
\end{center}
Now for Bob, the measurement Probability of getting $\ket{00}$ and 
$\ket{11}$ are: 
\begin{eqnarray*}
Tr(\ket{00}\bra{00}\rho') &=& \frac{1}{2}(1+\lambda^2) = \frac{1}{2}(2-2e^{-t/T_{1}}+e^{-2t/T_{1}}) \\
Tr(\ket{11}\bra{11}\rho') &=& \frac{1}{2}(1-\lambda)^2=\frac{1}{2}e^{-2t/T_{1}} \\
\end{eqnarray*}

At time $t=0$, the measurement probabilities of $\ket{00}$ and $\ket{11}$ are $\frac{1}{2}$ each, but as $t \rightarrow \infty$, the measurement probability of $\ket{00}$ approaches 1, and that of $\ket{11}$ decays to 0. However, that is not the only distortion created to the state. For $t > 0$, the entanglement is destroyed, and measurement yields the states $\ket{01}$ and $\ket{10}$ with probabilities:
\begin{eqnarray*}
Tr(\ket{01}\bra{01}\rho') &=& Tr(\ket{10}\bra{10}\rho') = \frac{1}{2} \lambda(1-\lambda)\\
&=& \frac{1}{2} e^{-t/T_{1}} (1-e^{-t/T_{1}})
\end{eqnarray*}

The measurement probability for both $\ket{01}$ and $\ket{10}$ is the same. When t=0, the measurement probability is 0 and as t $\rightarrow \infty$, the measurement probability is still 0. So, in order to get a better look at the behaviour of the expression, we are looking for a possible maxima or minima on the success probability of the protocol. The expression for the probability of success is:
\begin{eqnarray*}
\text{P(Success)} &=& 1 - P(\ket{01}) - P(\ket{10})\\
&=&(1-e^{-t/T_1} + e^{-2t/T_1})
\end{eqnarray*}

Differentiating the above expression of success probability we get:
\begin{center}
      $\frac{d}{dt}(1-e^{-t/T_1} + e^{-2t/T_1}) = \frac{1}{T_{1}}(e^{-t/T_{1}}-2e^{2t/T_{1}})$ 
\end{center}
and double differentiating the probability measurement we get:
\begin{center}
    $\frac{d^{2}}{dt^{2}}(1-e^{-t/T_1} + e^{-2t/T_1}) = \frac{1}{T_{1}^{2}}(-e^{-t/T_{1}}+4e^{-2t/T_{1}})$ 
\end{center}
Equating the differentiated expression to 0, we get:
\begin{center}
 $\frac{1}{T_{1}}(e^{-t/T_{1}}-2e^{2t/T_{1}})=0$\\
$\text{or, } t=T_{1}ln(2)$
\end{center}  

\begin{figure}[H]%
    \centering
    {{\includegraphics[width=7.5cm]{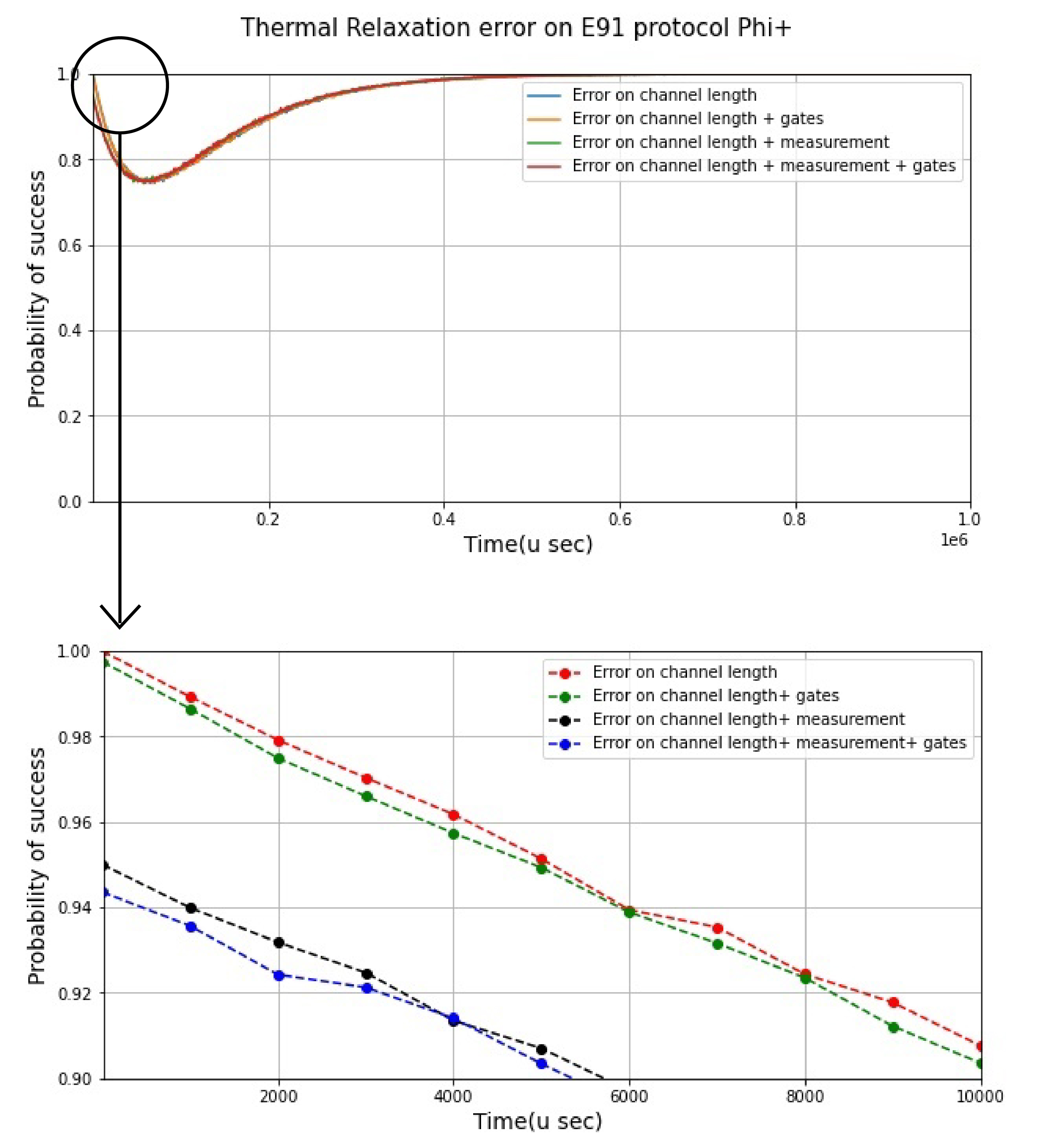} }}%
    \caption{ Probability of Success}%
    \label{fig:var4}%
\end{figure}

Putting the value of $t=T_{1}ln(2)$ in the doubly differentiated expression we get $\frac{1}{2T_{1}^{2}}$ , which is  greater than 0. Thus, we can say that the extremum point is a minima, which matches our result with the plot from the simulation. Also, the peak of measurement probability for $\ket{01}$ and $\ket{10}$ is at the same point of time (which is around 13130734.489 $nano$ sec for our $T_{1}$ value of 188610 $nano$ sec). The probability of success at time t=0 is 1 and as t $\rightarrow \infty$ the probability is still 1, and the success probability is minimum at $t=T_{1}ln(2)$ which is confirmed by Fig.~\ref{fig:var4}.

Now, it seems the probability of success in Fig.~\ref{fig:var4} approaches a minima after certain time and then again increases which is very counter-intuitive but looking at the individual graphs of $\ket{00}$ and $\ket{11}$, this is occurring due to system collapses to $\ket{00}$ which is increasing gradually. But, the problem with the success probability is that, the very definition of it is very biased. The definition completely overlooks the entanglement of the state and just focuses on the measurement outcome. From the very beginning, there is the small chance of getting $\ket{01}$ or $\ket{10}$, which can be observed from Fig.~\ref{fig:var5}.b, and thus, destroying the entanglement from the very start. Now from Eve's perspective, if she decides to guess $\ket{00}$ at the beginning, her probability of being correct would be $(1-e^{-t/T_{1}}+e^{-2t/T_{1}})$, i.e., at time t=0, her probability of correctly guessing would be $\frac{1}{2}$ and as time increases the value would approach to 1 again.
 The above analytical results are substantiated by Fig.~\ref{fig:var5} with the help of simulation. For this case, we also used the same value of T1 as our former simulation. For simulating the E91 protocol, we designed the circuit for $\ket{\phi^{+}}$ and subjected it under thermal relaxation error. The resulting graph for the measurement outcomes are as follows :  
\begin{figure}[H]%
    \centering
    \subfloat[\centering Change of Probability of Bob measuring $\ket{00}$ and $\ket{11}$ ]{{\includegraphics[width=7.5cm]{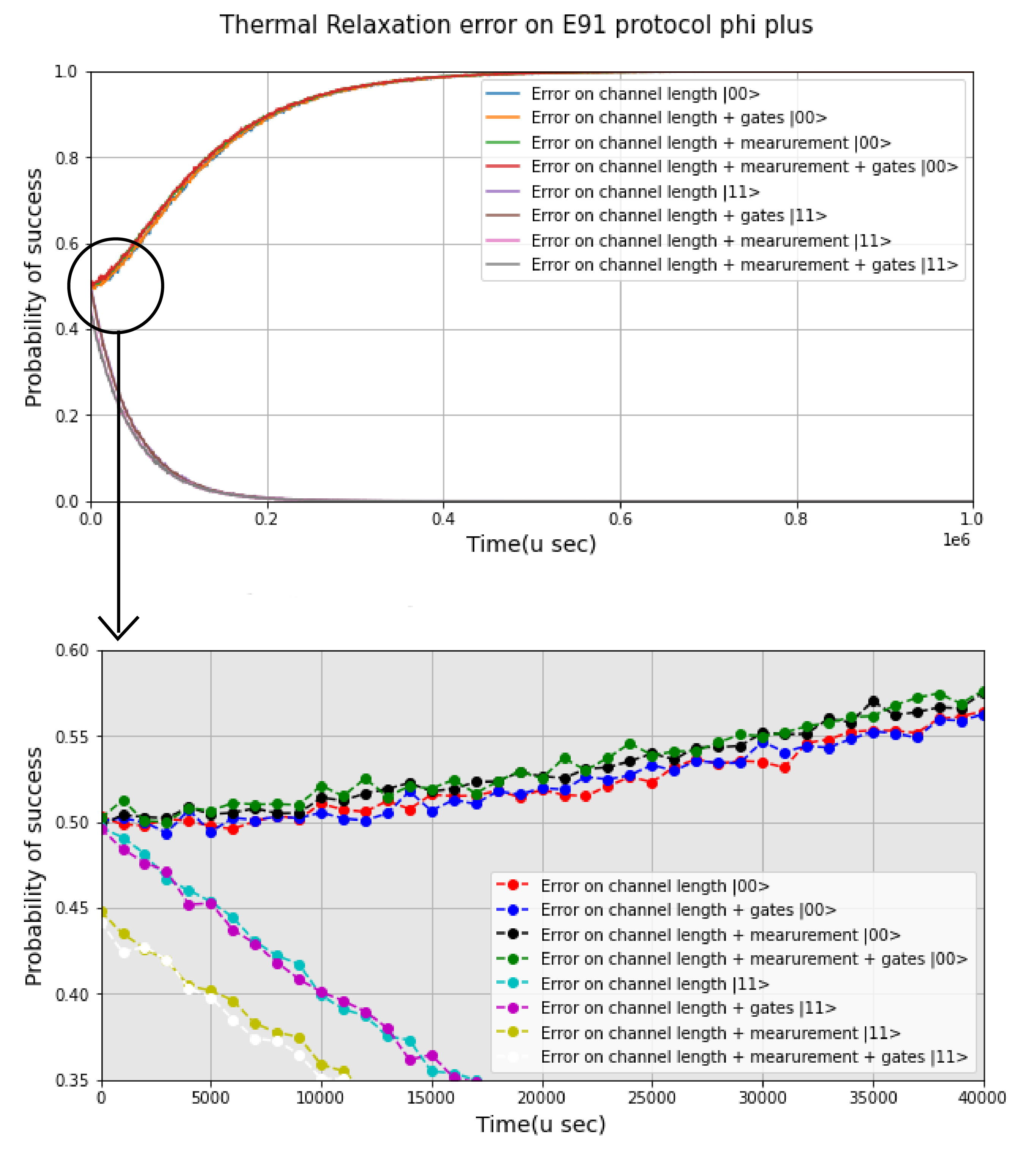} }}%
    \centering
    \qquad
    \subfloat[\centering Change of Probability of Bob measuring $\ket{01}$ and $\ket{10}$ ]{{\includegraphics[width=7.5cm]{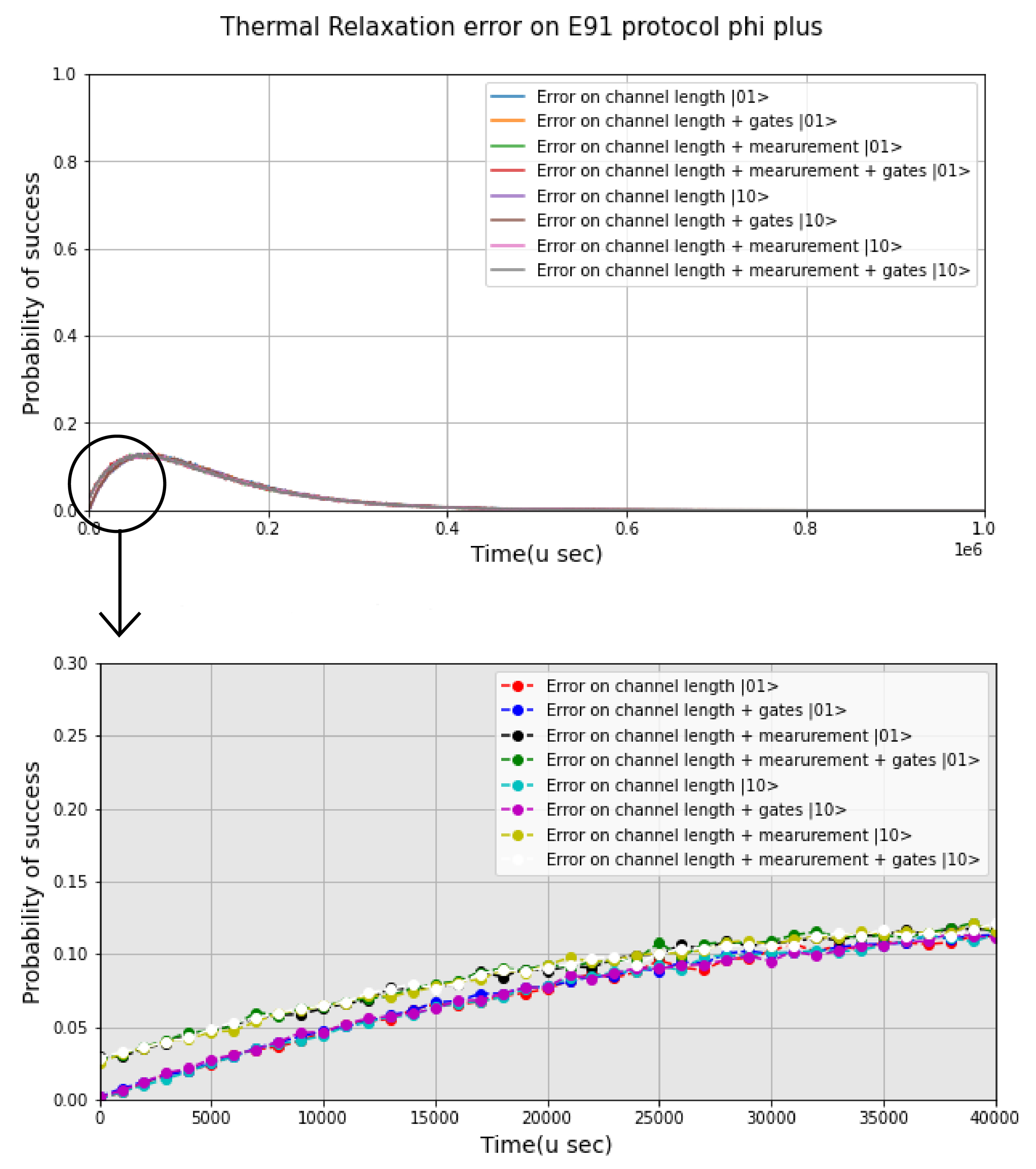} }}%
    \caption{Measurement probability of outcome states in E91 using $\ket{\phi^{+}}$ under Thermal Relaxation Error }%
    \label{fig:var5}%
\end{figure}

\subsection{E91 protocol with $\ket{\psi^{+}}$}
Now, Let Alice and Bob are using only :
  \begin{center}
  
$\ket{\psi^{+}} = \frac{1}{\sqrt{2}}(\ket{01}+\ket{10})$\\
  
$\therefore \rho = \ket{\psi^{+}}\bra{\psi^{+}} = \frac{1}{2}\begin{bmatrix}
0 & 0 & 0 & 0\\
0 & 1 & 1 & 0\\
0 & 1 & 1 & 0\\
0 & 0 & 0 & 0
\end{bmatrix}$\\  
$\therefore \rho' =\frac{1}{2}\begin{bmatrix}
2\lambda & 0 & 0 & 0\\
0 & (1-\lambda) & (1-\lambda) & 0\\
0 & (1-\lambda) & (1-\lambda) & 0\\
0 & 0 & 0 & 0
\end{bmatrix} $\\\
  \end{center}
  
Measurement probability of $\ket{01}$ and $\ket{10}$ are:
\begin{eqnarray*}
    Tr(\ket{01}\bra{01} \rho')&=&\frac{1}{2}(1-\lambda)=\frac{1}{2}e^{-t/T_{1}}\\
    Tr(\ket{10}\bra{10} \rho')&=&\frac{1}{2}(1-\lambda)=\frac{1}{2}e^{-t/T_{1}}
\end{eqnarray*}

As the measurement probabilities of $\ket{01}$ and $\ket{10}$ decrease, the other state that takes their place is $\ket{00}$. The measurement probability of $\ket{00}$ is : $Tr(\ket{00}\bra{00} \rho')=\lambda=(1-e^{-t/T_{1}})$

\begin{figure}[H]%
    \centering
    \subfloat[\centering Change of Probability of Bob measuring $\ket{01}$ and $\ket{10}$ ]{{\includegraphics[width=7.5cm]{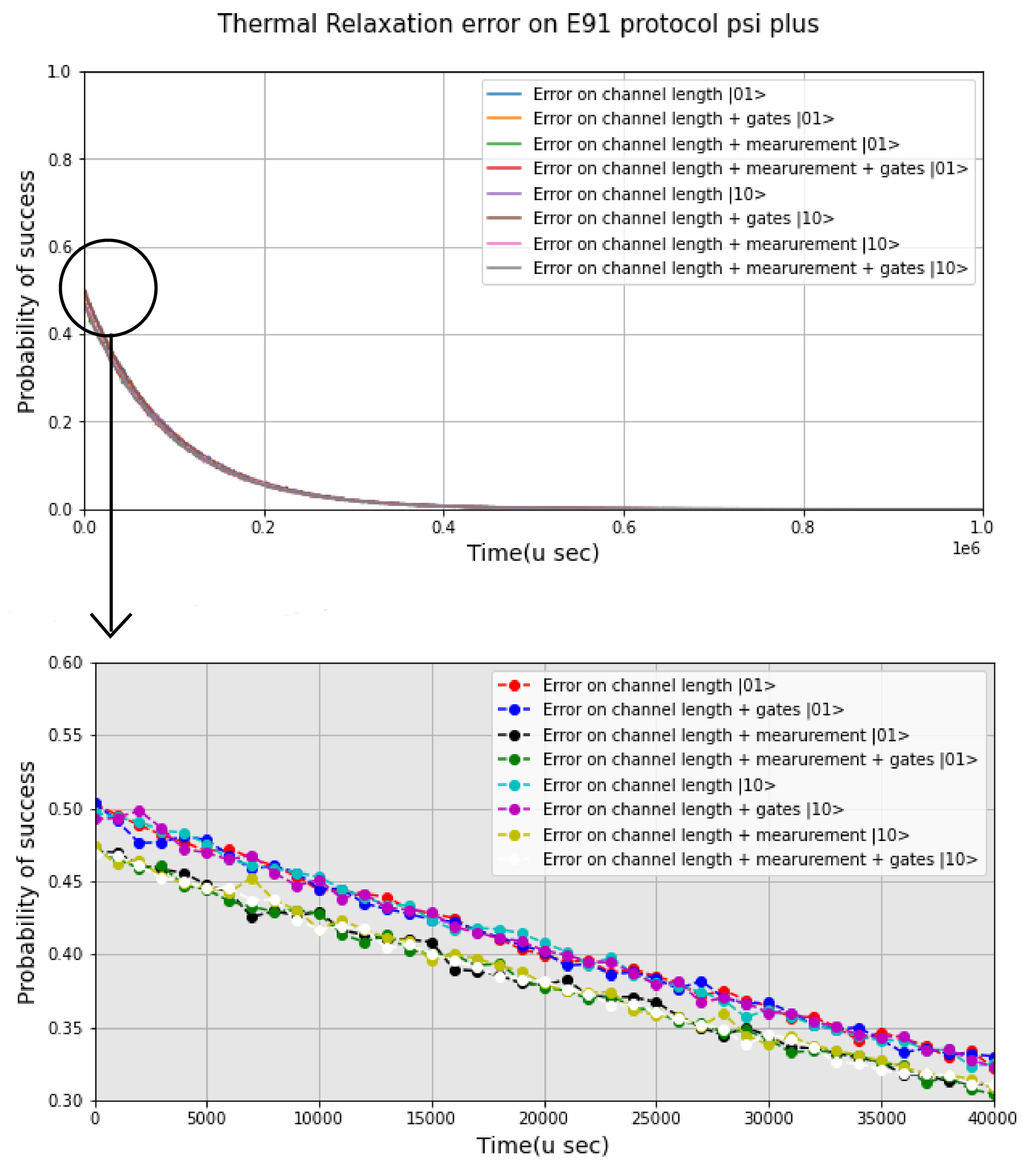} }}%
    \centering
    \qquad
    \subfloat[\centering Change of Probability of Bob measuring $\ket{00}$ ]{{\includegraphics[width=7.5cm]{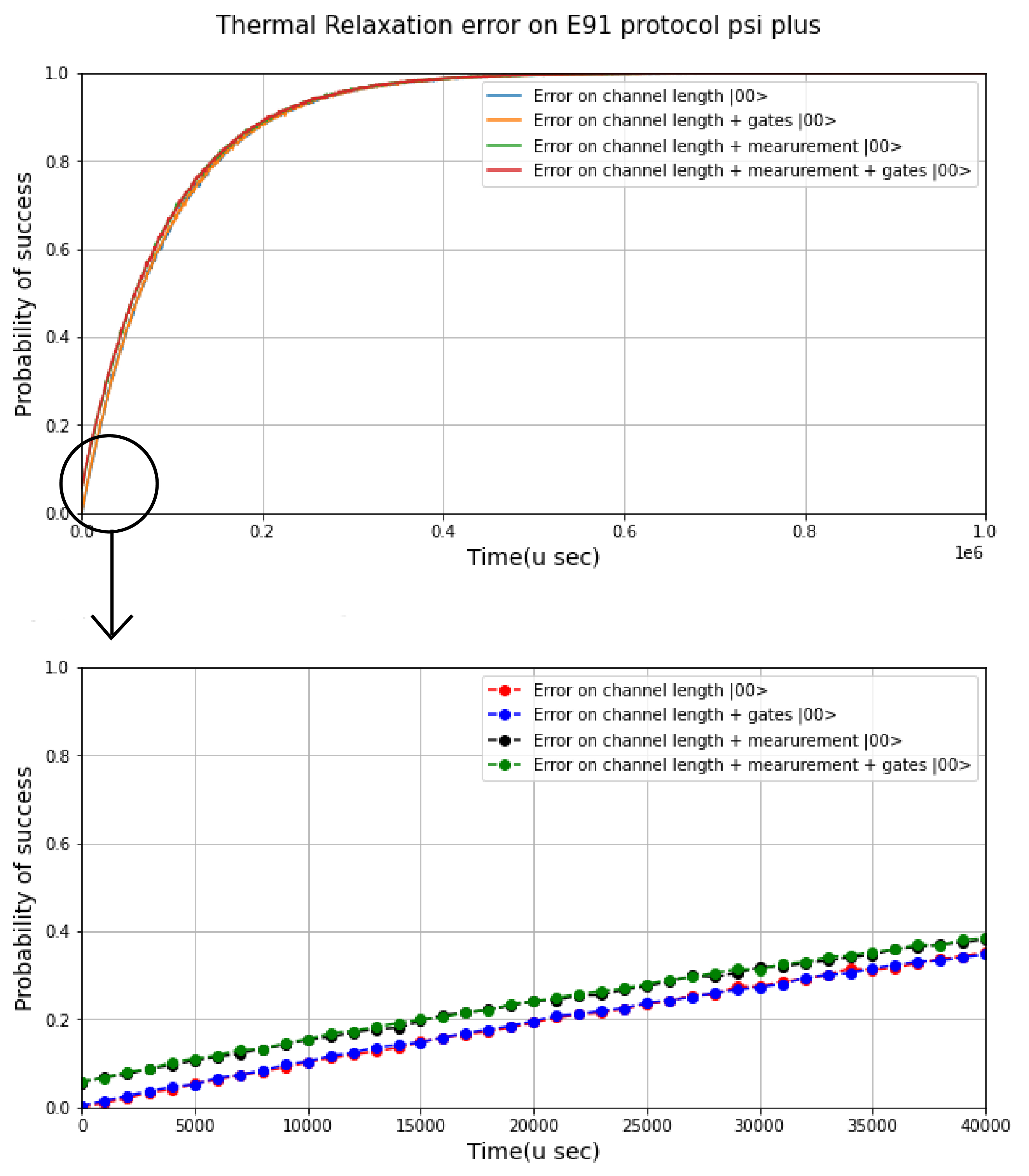} }}%
    \caption{Measurement probability of outcome states in E91 using $\ket{\psi^{+}}$ under Thermal Relaxation Error }%
    \label{fig:var6}%
\end{figure}
So, both the states decay at an exponential rate and collapses to $\ket{00}$ as time increases. At time t=0, the measurement probability of both $\ket{01}$ and $\ket{10}$ is half, and as t$\rightarrow \infty$, the probability decreases to 0. In case of $\ket{00}$, it starts with 0 but with increasing time, the probability approaches 1. These conclusions are complemented by our simulation results in Fig.~\ref{fig:var6}, which is simulated under same conditions as previous one.

Probability of getting $\ket{01} and\ket{10}$ are- \\* 
\[P(\ket{01}) = P(\ket{10}) = \frac{1}{2}e^{-t/T_{1}} \] 
\[\text{P(Success)} = \frac{1}{2} e^{-t/T_{1}} \times 2 =e^{-t/T_1}\]

\begin{figure}[H]%
    \centering
    {{\includegraphics[width=7.5cm]{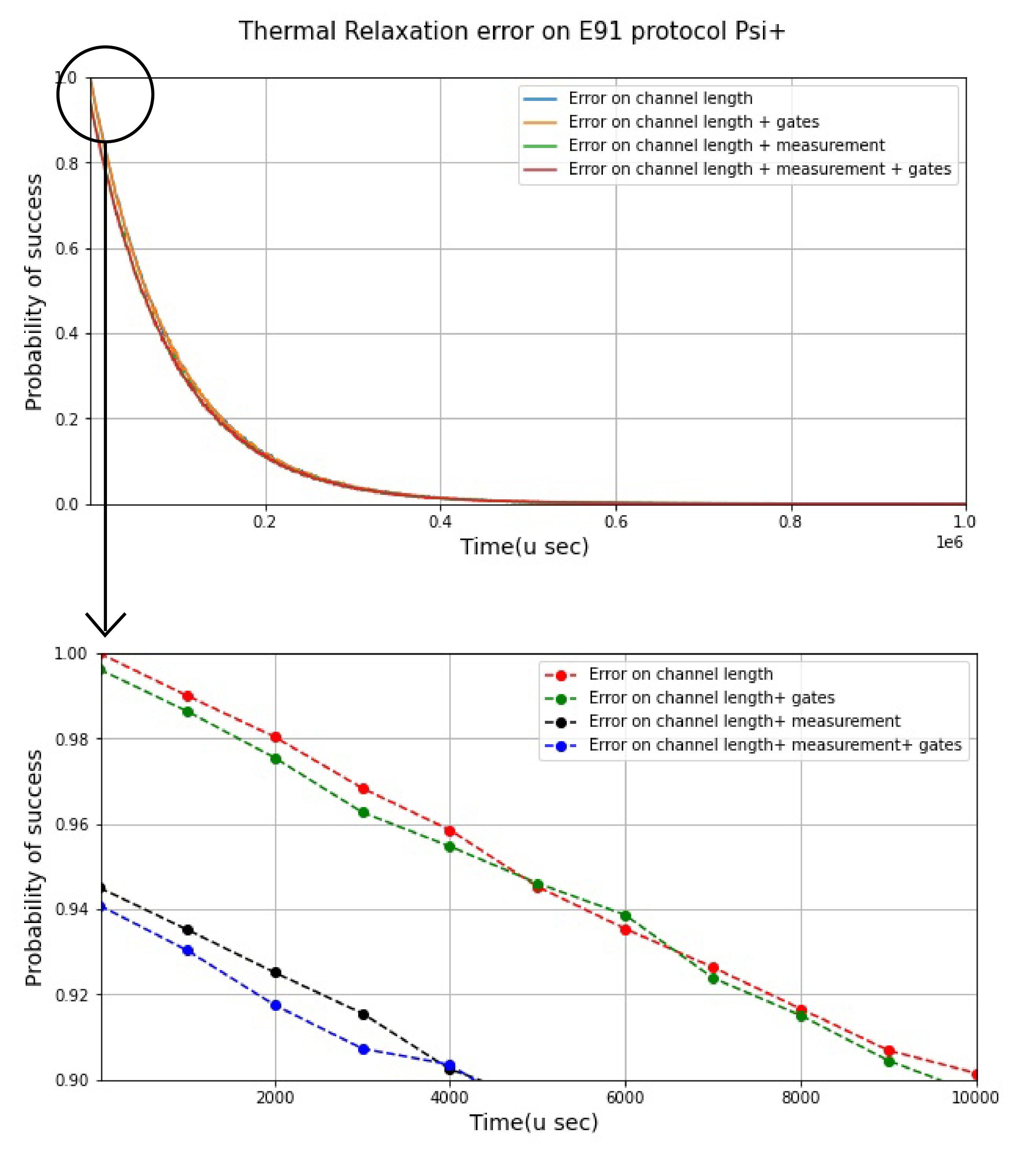} }}%
    \caption{ Probability of Success}%
    \label{fig:var7}%
\end{figure}

So, when Alice and Bob chose to work with $\ket{01}$ and $\ket{10}$ states, they collapse to $\ket{00}$ again. At the beginning, the chance of getting 00 is small, but this increases gradually, thus destroying entanglement again at the very beginning. The resultant graph from the simulation depicts the same conclusion.\\

The entanglement property of Bell states gives Alice and Bob substantial advantage over Eve. Eve will have to interact with the bits to know the states, notifying Alice and Bob about the breach in the protocol. But to get the same conclusion in BB84, Alice has to use half of the bits measured in same basis by Alice and Bob. But considering the protocols in the presence of the Thermal relaxation error, BB84 protocol is the better choice compared to E91 protocol. The reason behind it is that, use of either $\ket{\phi^{+}}$ or $\ket{\psi^{+}}$ leads to destruction of the entanglement from the very beginning, which is evident from the simulation plots. But the list of reasons does not end there. Assuming Eve knows about the error in the channel, she will be able to entangle her own state to the state exchanged between Alice and Bob, which basically will result into a failure of the secret information exchange between Alice and Bob since Eve will be aware about their measurement. We should not be influenced by the success probability of E91 protocol with $\ket{\phi^{+}}$, due to the fact that the probability of success for E91 protocol is very deceptive, since it does not reflect on the destruction of the entanglement of protocol and is merely just a probabilistic outcome of the states that Alice and Bob are supposed to get measured in same basis. For E91 protocol with $\ket{\psi^{+}}$ state, with increasing time, Alice and Bob have higher chances of getting $\ket{00}$ state, which is not meant to be there at all and also the entanglement breaks at the very beginning. BB84, in this case, does not rely on entanglement. So, the only thing we face is the decay of the excited states.  But the resulting situation gives Eve a substantial advantage in guessing the states which we expand in the next section.\\

So, to conclude, in presence of thermal relaxation error, E91 performs poorly. This is because, in both the cases, the error destroys the entanglement at the very beginning, which is the very essence of the E91 protocol. On the other hand, BB84 protocol has sufficient scopes of improvement in performance under thermal relaxation error with a minor modification, that we have discussed in the next section.

\section{Guessing power of Eve and proposed modified BB84 protocol}
In this section, we consider the presence of an eavesdropper Eve, who-
\begin{itemize}
    \item has complete knowledge about the noise in the channel, and the duration of the protocol,
    \item \emph{makes a guess} based on that knowledge.
\end{itemize}

This poses a significant restriction on the capability of Eve. However, for this preliminary study, we restrict the capacity of Eve to \emph{guessing} only. There may be multitude of techniques that Eve can exploit with the noise information, which we postpone for future studies. We show in this article that with the knowledge of channel noise, even an Eve who simply guesses, can take an advantage. For BB84 protocol without noise, it is statistically possible to detect the presence of Eve if Alice and Bob randomly discuss a few measurement outcomes \cite{bb84}. However, if Eve makes a guess regarding the outcome, then the parties has no information about her presence, and she can hide herself behind the noise in the system.

In the noiseless scenario, since each of the four state is prepared randomly with equal probability, Eve can guess correctly with a probability of $\frac{1}{4}$. In the following, we show that, when equipped with the knowledge of the noise in the channel, Eve can guess with probability $\frac{1}{4} + \epsilon$, for $\epsilon > 0$. Finally, we identify the reason for higher guessing probability of Eve, and suggest a simple modification of the BB84 protocol which is equivalent to the original protocol when there is no noise, and yet restricts Eve's guessing to a probability of $\frac{1}{4}$ in a thermal relaxation channel.

\subsection{Better guessing on BB84 protocol with noise information}
We assume that Eve has perfect information of the channel noise, i.e., she knows (i) it is a thermal relaxation channel, (ii) the $T_1$ time of the device used, and (iii) the time $t$ of the execution of the protocol. Equipped with this knowledge, she guesses the outcome at Bob's end. Assuming an ideal condition where there is no error in the channel, each state $\{\ket{0}, \ket{1}, \ket{+}, \ket{-}\}$ is sent with probability of $\frac{1}{4}$. If Eve guesses, she is successful with probability $\frac{1}{4}$ only. We show here that with the information of channel noise, Eve can design a strategy to guess with probability $\frac{1}{4} + \epsilon$, for $\epsilon > 0$.

\textbf{A strategy for Eve:}
Let Eve make a strategy where she guesses the basis $X$ and $Z$ with equal probability. When she chooses the basis $X$, she guesses the state to be $\ket{+}$ or $\ket{-}$ with equal probability. But when she chooses the basis $Z$, she always guesses the state to be $\ket{0}$.



\textbf{Probability of success for Eve:} The probability that Eve guesses correctly is
\begin{center}
    $\sum_{x \in \{0,1,+,-\}}$ Prob that the state is $x$ $\times$ Prob Eve guesses $x$
\end{center}

When there is no noise, each term in the above summation is $\frac{1}{16}$, and so the total probability that Eve guesses correctly is $\frac{1}{4}$.

Now, even in the presence of noise, we have already shown that both $\ket{+}$ and $\ket{-}$ decay at the same rate. Therefore, the probability distribution of the outcome of these two states remain the same, and Eve does not gain any advantage when choosing one over the other. So, we can safely say that for $+, -$, each term in the summation is still $\frac{1}{16}$.

But the strategy of Eve is that she always chooses $\ket{0}$ for $Z$ basis. Therefore, if the outcome is $\ket{1}$, she is always wrong. But,
\begin{center}
    Prob that the state is $\ket{0}$ $\times$ Prob Eve guesses $\ket{0}$\\
    = $(\frac{1}{4}+\{\frac{1}{4}(1-e^{(-t/T_1)})\} \times \frac{1}{2}$\\
    = $\frac{1}{8} + \epsilon$
\end{center}

where the term $\frac{1}{4}(1-e^{(-t/T_1)} = 2\epsilon$ is the probability of measuring as state $\ket{1}$ as $\ket{0}$ after time $t$ due to thermal relaxation. We note that $\epsilon \geq 0$ for $t \geq 0$, and for $t > 0$, $\epsilon > 0$. Considering the correct probabilities of guessing when the states are $\ket{+}$ or $\ket{-}$, with this strategy, Eve can correctly guess the state with probability $\frac{1}{4} + \epsilon$, which is greater than the noiseless scenario, and still remain perfectly hidden from the parties.

In the next subsection we propose a simple modification to the BB84 protocol, particularly for the thermal relaxation channel, that strips Eve of this advantage in guessing.

\subsection{Proposed modification to BB84 protocol}
In the previous subsection, we showed a strategy that Eve can follow to guess better under thermal relaxation error. Now, we propose a modification to the original BB84 protocol so that Eve fails to attain any upper hand in guessing even when equipped with the full information of the thermal relaxation channel.

In the modified protocol, instead of preparing the states in the eigenstates of $X$ and $Z$ basis, Alice prepares them in the eigenstates of $X$ and $Y$ basis. $\ket{+i}$ and $\ket{-i}$ states are the eigenstates of Y basis. Alice would prepare $\ket{+}$, $\ket{-}$, $\ket{+i}$, $\ket{-i}$ with equal probability. The rest of the protocol remains the same. This variation does not affect the security of the original protocol, since this can now be thought of as the BB84 protocol, but in a rotated space. However, we show that this simple change stops Eve from having any advantage in guessing. This is because the decay rate of all the four states are now same.

Consider the following:

\begin{center}
$ \ket{+i}= \frac{1}{\sqrt{2}}(\ket{0}+ i\ket{1}),\ket{-i}= \frac{1}{\sqrt{2}}(\ket{0}- i\ket{1})$ \\
 $\rho_{+i}' =\frac{1}{2}\begin{bmatrix}
1+\lambda & -i(\sqrt{1-\lambda})\\
i(\sqrt{1-\lambda}) & 1- \lambda
\end{bmatrix}$\\

 $\rho_{-i}' =\frac{1}{2}\begin{bmatrix}
1+\lambda & i(\sqrt{1-\lambda})\\
-i(\sqrt{1-\lambda}) & 1- \lambda
\end{bmatrix}$ 
\end{center}
The measurement probability of the states are:
\begin{center}
$Tr(\ket{+i}\bra{+i} \rho_{+i}')=\frac{1}{2}(1+ \sqrt{1-\lambda})=\frac{1}{2}(1+e^{-t/2T_{1}})$

$Tr(\ket{-i}\bra{-i} \rho_{-i}')=\frac{1}{2}(1+ \sqrt{1-\lambda})=\frac{1}{2}(1+e^{-t/2T_{1}})$
\end{center}

So, the probability of decay of both the states is same. Thus the probability of Bob measuring the same state as Alice is:
\begin{eqnarray*}
\text{P(B=$+i$\textbar A=$+i$)}&=&\text{P(B=$-i$\textbar A=$-i$)}\\
&=&\frac{1}{2}(1+e^{-t/2T_{1}})
\end{eqnarray*}

So in this modified protocol, all the states decay at the same rate. Therefore, it is not possible for Eve to make a better guess even if she has all the information about the duration of the protocol, and the characteristics of the channel. So the probability that Eve can guess correctly is still $\frac{1}{4}$.

\begin{figure}[htp]%
    \centering
    {{\includegraphics[width=7.5cm]{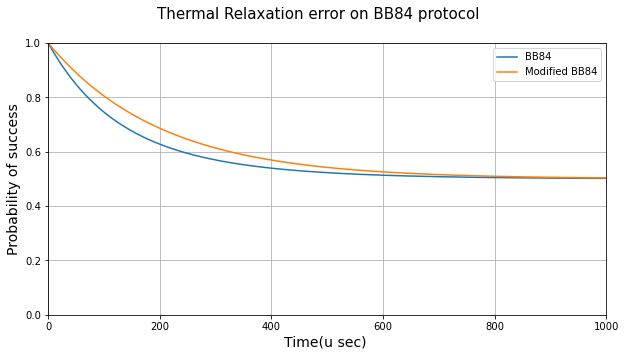} }}%
    \caption{Total Success probability Comparison of BB84 protocol and Modified BB84 protocol}%
    \label{fig:var8}%
\end{figure}

 
\section{Conclusion}
 In case of BB84 protocol, it is shown that all the prepared states decay except the 0 state, but what varies is the slope of the curve for the states i.e., the decay rate. The decay rate of 1 bit is much higher than the Hadamard states prepared by Alice. Using the Z basis states gives Eve a substantial advantage of guessing Bob’s measured bit as with $t$ approaching infinity, the probability of measuring 0 approaches to $\frac{1}{2}$. So with increasing time, if Eve guesses 0, her chances of correctly guessing Bob's measurement increases. But in case of $X$ bases, the probability of decay is same for both bits and the decay rate is also much slower than that of 1 bit. That leads to a possible modification for this protocol to use states in Hadamard bases and Y basis.

The use of $X$ and $Y$ basis states saves us from increasing Eve’s probability of guessing Bob's measurement, and without any better guess, Eve has to interact with the prepared bits. But for the E91 protocol, it is clearly observed that the Thermal relaxation error completely destroys the entanglement of the prepared state which is the very essence of the protocol, thus making entanglement based protocol highly unlikely to work under this kind of error.

In this work, we restricted Bob's ability to guessing only and proposed a modification focusing on it. A future prospect of this research can be to consider other eavesdropping strategies for Eve under similar conditions.

\bibliographystyle{unsrt}
\bibliography{reference.bib}

\end{document}